\documentclass[a4paper, amsfonts, amssymb, amsmath, reprint, showkeys, nofootinbib, twoside, onecolumn,notitlepage]{revtex4-1}

\def\apj{{ApJ}}                 
\def\aap{{A\&A}}                

\def\mnras{{MNRAS}}             
\def\prd{{Phys.~Rev.~D}}        
\def\pre{{Phys.~Rev.~E}}        
\usepackage{amsmath,amstext}
\usepackage[T1]{fontenc}
\usepackage{amssymb}
\input{epsf}
\usepackage{graphicx}
\usepackage{ae,aecompl}

\usepackage{hyperref}
\usepackage{amsmath}
\usepackage{amssymb}
\usepackage{mathtools}
\usepackage{bm}
\usepackage{cleveref}
\usepackage{tensor}
\usepackage{braket}
\usepackage{enumitem}
\usepackage{mhchem}

\usepackage{color}

\newcommand{\spc}{\quad \quad \quad}

\newcommand{\riga}{\textcolor{white}{-}}

\def\be{\begin{equation}}
\def\ee{\end{equation}}
\def\beq{\begin{eqnarray}}
\def\eeq{\end{eqnarray}}

\begin{document}
\title{Applying the Gibbs stability criterion to relativistic hydrodynamics}
\author{L.~Gavassino}
\affiliation{Nicolaus Copernicus Astronomical Center, Polish Academy of Sciences, ul. Bartycka 18, 00-716 Warsaw, Poland}

\begin{abstract}
The stability of the equilibrium state is one of the crucial tests a hydrodynamic theory needs to pass. A widespread technique to study this property consists of searching for a Lyapunov function of the linearised theory, in the form of a quadratic energy-like functional. For relativistic fluids, the explicit expression of such a functional is often found by guessing and lacks a clear physical interpretation. We present a quick, rigorous and systematic technique for constructing the functional of a generic relativistic fluid theory, based on the maximum entropy principle. The method gives the expected result in those cases in which the functional was already known. For the method to be applicable, there must be an entropy current with non-negative four-divergence. This result is an important step towards a definitive resolution of the major open problems connected with relativistic dissipation.
\end{abstract}

\maketitle

\section{Introduction} 

Recent years have seen an explosion of new dissipative hydrodynamic theories, as fluid descriptions are applied to different fields, ranging from heavy ion collisions  \cite{FlorkowskiReview2018}, to neutron star physics \cite{AnderssonMultifluidPerspective2021} and cosmology \cite{Maartens1995}. The demand for new theories comes from the inadequacy of simple fluids to account for the complexity of real systems.
For example, current theories for viscosity \cite{Israel_Stewart_1979,Liu1986} fail to describe the initial transient of strongly interacting quantum field theories \cite{Denicol_Relaxation_2011, Heller2014}. Furthermore, cold neutron-star matter is a superfluid-normal mixture, which requires multi-fluid modelling \cite{carter1991,Carter_starting_point,langlois98,Gusakov2007,Termo}. Even less exotic astrophysical systems (such as accretion disks and jets) cannot accurately be described using simple fluids, due to the presence of a magnetic field, a radiation field and two-temperature effects \cite{FernandezGRMHD2019,Sadowski2013,Mosibroska2016}. Combining these features with causal dissipation leads to completely new theories, e.g. \cite{AnileRadiazion1992}. Finally, hot dense matter in supernovae and neutron-star mergers is a reacting mixture, with reaction time-scales comparable with the hydrodynamic time-scale \cite{Burrows1986,Alford2020,Nedora2021}. This requires us to revise our understanding of causal bulk viscosity \cite{BulkGavassino}.  

As more and more complex theories are proposed, it is of central importance to be able to predict if the theory that one is building is truly dissipative (i.e. if the fluid exhibits a tendency to evolve towards thermodynamic equilibrium) or if the non-equilibrium degrees of freedom undergo a non-physical spontaneous explosion, as in the case of the theories of Eckart and Landau-Lifshitz \cite{Hiscock_Insatibility_first_order,Hishcock1988}. This criterion of \textit{stability of the equilibrium}, according to which states that are initially close to global \textit{thermodynamic} equilibrium  \cite{VANWEERT1982,Israel_2009_inbook,BecattiniBeta2016,Salazar2020,GavassinoTermometri} remain close to it, constitutes the most fundamental reliability test of a dissipative theory \cite{GavassinoFronntiers2021}. Unfortunately, verifying this property with the current techniques is usually complicated and the physical interpretation of the stability conditions is often not transparent \cite{Geroch95,LindblomRelaxation1996,Kost2000,Straughan2004}. In fact, the calculation strongly depends on the details of the hydrodynamic equations: adding a new coupling or slightly modifying the physical setting might force one to start over the whole stability analysis \cite{OlsonLifsh1990,Straughan2004,Brito2020}. More importantly, one would like to be able to test the stability of any possible thermodynamic equilibrium state (at rest or in motion, rotating or non-rotating, with or without a strong gravitational field) at once, while often (when the theory becomes too complicated) the calculation is specialised to homogeneous equilibria in a Minkowski background \cite{VanBir2008,Stricker2019,Brito2020,Kovtun2019,BemficaDNDefinitivo2020,Lopez11}.

On the other hand, the theory of \textit{thermodynamic} stability has a long history, which goes back to Gibbs \cite{PrigoginebookModernThermodynamics2014}. The idea of Gibbs was simple: since the entropy cannot decrease, the equilibrium state of an isolated system is stable if any (physically allowed) perturbation results in a decrease in entropy. In other words, the entropy should be maximum in equilibrium, to ensure Lyapunov stability \cite{Prigogine1978}. Here, we apply this principle to relativistic hydrodynamics, presenting a technique to build, directly from the constitutive relations of a generic fluid, a quadratic Lyapunov functional, whose positive-definitiveness implies stability. Below we outline the methodology and we give a couple of examples and applications. We adopt the signature $(-,+,+,+)$ and work in units $c=k_B=1$.

\section{The stability criterion}

It is crucial for our method that we can associate to the fluid a symmetric stress-energy tensor $T^{ab}$ and an entropy current $s^a$ which obey the conditions
\begin{equation}\label{necessary}
\nabla_a T^{ab}=0 \quad \quad \quad \nabla_a s^a \geq 0
\end{equation} 
as \textit{exact} mathematical constraints. The remaining details of the field equations (such as the exact value of the entropy production rate) are irrelevant and do not play any role in the method, provided that \eqref{necessary} are respected. Here we assume, for illustrative purposes, that there is a single conserved current $N^a$, such that $\nabla_a N^a=0$, but the method can be straightforwardly generalised to fluids with an arbitrary number of conserved currents.
We assume that the fluid is immersed in a test spacetime (which plays the role of a fixed background), having one and only one Killing vector field $K^a$, which is everywhere time-like future-directed. If the fluid has a finite spatial extension, then, assigned a space-like Cauchy 3D-surface $\Sigma$, the three integrals
\begin{equation}\label{due}
\{ \, N, \, U, \, S \, \} = \int_\Sigma \{ \, -N^a, \, T^{ab}K_b, \, -s^a \, \} \, d\Sigma_a \, 
\end{equation}
are finite and represent the total particle number, energy and entropy of the fluid. Given the aforementioned assumptions, $N$ and $U$ are conserved (i.e. they do not depend on $\Sigma$), while 
\begin{equation}\label{Entrppi}
S \,[\Sigma'] \geq S \,[\Sigma]
\end{equation}
whenever $\Sigma'$ is in the future of $\Sigma$. We also need to have a selection of the macroscopic fields $\varphi_i$ which carry information about the local state of the fluid (e.g., for the perfect fluid one may take the fluid velocity, the temperature and the chemical potential) and the constitutive relations:
\begin{equation}\label{constutiveingeneral}
T^{ab}=T^{ab}[\varphi_i] \quad \quad s^{a}=s^{a}[\varphi_i] \quad \quad N^{a}=N^{a}[\varphi_i].
\end{equation}

The method works as follows: we consider two solutions of the (in principle unknown) hydrodynamic equations, which are close to each other,
\begin{equation}
\varphi_i  \spc \text{and}  \spc \varphi_i + \delta \varphi_i \, ,
\end{equation}
and we define the variation of any observable $\mathcal{A}$ as the \textit{exact} difference \footnote{We avoid notations like ``$\, \delta \,$'' and ``$\, \delta^2 \,$'' for first and second order variations. We, instead, introduce $\delta \mathcal{A}$ as an exact difference, which is later approximated according to the need.}
\begin{equation}
\delta \mathcal{A} := \mathcal{A}[\varphi_i+\delta \varphi_i] - \mathcal{A}[\varphi_i].
\end{equation}
The configuration $\varphi_i$ is our candidate global thermodynamic equilibrium state, while $\delta \varphi_i$ is a deviation from equilibrium which should decay to zero for large times (if the theory is dissipative and stable). For this to be possible, we must impose that the integrals of motion $U$ and $N$ have exactly the same value in the two states, namely
\begin{equation}\label{deltaNunz}
\delta N = \delta U =0,
\end{equation}  
otherwise $\varphi_i + \delta \varphi_i$ would asymptotically relax to an equilibrium state which is different from $\varphi_i$ \footnote{If we interpret $\varphi_i + \delta \varphi_i$ as the state just after an external kick has been impressed, the constraint \eqref{deltaNunz} implies that $\varphi_i$ should not be interpreted as the state \textit{before} the kick (because the kick might not conserve $U$). Rather, $\varphi_i$ is the \textit{end} state of the evolution of $\varphi_i + \delta \varphi_i$, which is reached once thermodynamic equilibrium is re-established}. 
If there are additional constants of motion, like e.g. a superfluid winding number \cite{Termo}, these need to be treated on the same footing as $N$ and $U$. 

Now we only need to take two steps:
\begin{enumerate}
\item[i -] We truncate all the differences $\delta \mathcal{A}$ to the first order in $\delta \varphi_i$ and we impose the stationarity condition $\delta S=0$ for any possible choice of $\delta \varphi_i$ compatible with \eqref{deltaNunz}. This procedure \textit{defines} the thermodynamic equilibrium state $\varphi_i$ and identifies it completely. At this stage (and only for step i), one may deal with the constraint \eqref{deltaNunz} using some Lagrange multipliers $\alpha$ and $\beta$, rediscovering the covariant Gibbs relation \cite{Israel_Stewart_1979,VANWEERT1982,Israel_2009_inbook} 
\begin{equation}
\int_\Sigma \big(\delta s^a + \alpha \delta N^a + \beta K_b \delta T^{ab} \big) \,d\Sigma_a =0 \spc (\text{to first order}) \, .
\end{equation} 
\item[ii -] We go up in the truncation of all the quantities $\delta \mathcal{A}$ to the second order in $\delta \varphi_i$ and we study the sign of $\delta S$. Using the results of the previous step and recalling \eqref{deltaNunz}, we know that the first-order contribution vanishes, so that it is always possible to rewrite $E:=-\delta S$ as a quadratic functional in $\delta \varphi_i$. The Gibbs stability criterion requires us to impose its positive definiteness. 
\end{enumerate}
Now, since in equilibrium the entropy is conserved (it cannot increase further once it is maximal), the inequality \eqref{Entrppi} implies that $E$ cannot increase with time (namely, $E \,[\Sigma'] \leq E \,[\Sigma]$ for $\Sigma'$ future of $\Sigma$). This, combined with the requirement that $E > 0$ whenever $\delta \varphi_i \neq 0$, is a sufficient condition of Lyapunov stability (more precisely, perturbations have a bounded square integral norm \cite{Hishcock1983,GerochLindblom1990}). 

Furthermore, Hiscock and Lindblom proved (see Proposition B of the Appendix of \cite{Hishcock1983}), with an argument that can be applied to every theory consistent with \eqref{necessary}, that $E>0$ is also a necessary condition of stability. The intuitive idea is that, for $E=-\delta S$ to approach a finite value for large times, $\nabla_a \delta s^a$ must converge to zero. Thus, all solutions of stable dissipative theories asymptotically converge to solutions of non-dissipative theories. However, the physical properties of non-dissipative theories are determined by the equilibrium equation of state, which (if computed from statistical mechanics) gives positive $-\delta S$ by construction. As $E[\text{initial}]\geq E[\text{final}]$, the positive definiteness of $E=-\delta S$ follows.

Before moving to the concrete examples, let us make a further comment about the constraints \eqref{deltaNunz}. In all the examples that follow (both in the main text and in the supplementary material), we deal with all constraints in an exact way, in the sense that \eqref{deltaNunz} is only used to perform exact cancellations, which would remain valid also at higher orders than the second. However, often one may find it more convenient to work with unconstrained variations. In Supplementary Material: Part 1, we show how to convert the maximum entropy principle at fixed energy and particle number into the minimum grand-potential principle, with completely free variations. This can, sometimes, make calculations easier (the final result is, of course, the same).

\section{Examples}

\begin{figure}
\begin{center}
\includegraphics[width=0.5\textwidth]{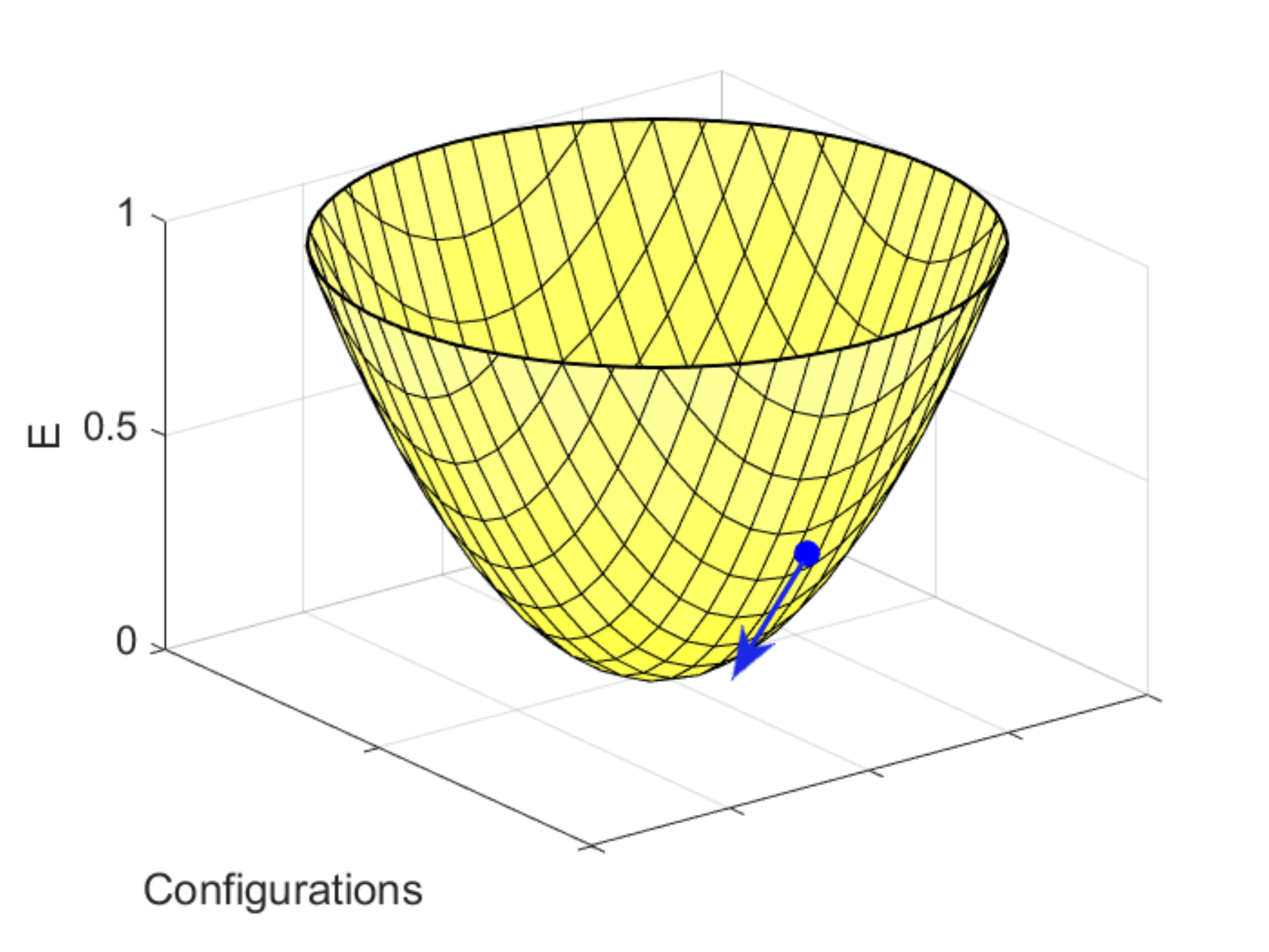}%
	\includegraphics[width=0.5\textwidth]{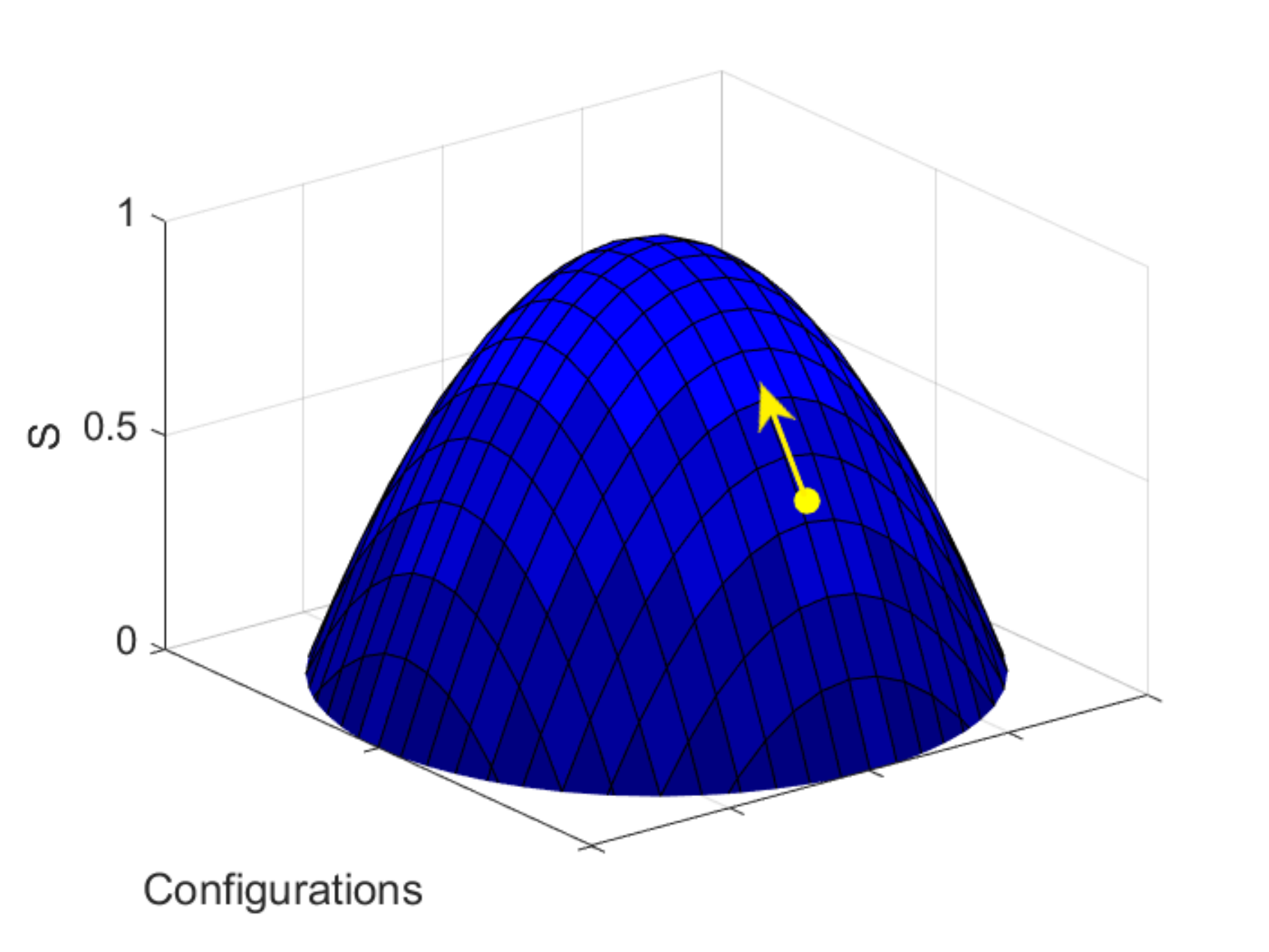}
	\caption{Two complementary views of the stability problem. Hydrodynamic view (left panel): the perturbation has a sort of ``energy functional'' $E \geq 0$, which decreases with time and should eventually converge to zero. Thermodynamic view (right panel): the perturbation reduces the entropy $S$; dissipation makes $S$ grow again to reach $S_{\text{eq}}$. The two pictures are connected by the equation  $E=S_{\text{eq}}-S$. The horizontal axes represent the abstract configuration space of the fluid, where each point is a global choice of $\delta \varphi_i$. The maximum of the entropy ($E=0$) is the equilibrium state ($\delta \varphi_i=0$).}
	\label{fig:fig}
	\end{center}
\end{figure}

To illustrate how the method works in practice, we consider the simplest possible causal theory for dissipation: the divergence-type theory \cite{Liu1986}. Adopting the notation of \citet{GerochLindblom1990}, the theory is built using three tensor fields, $\zeta_A = (\zeta,\zeta_a,\zeta_{ab})$, and postulates that there is a generating function $\chi =\chi (\zeta_A)$ such that
\begin{equation}\label{paritacchiuhc}
N^{aA} = \dfrac{\partial^2 \chi}{ \partial \zeta_a \partial \zeta_A} ,
\end{equation}
where we have grouped the three fluxes of the theory using the notation $N^{aA} =(N^a,T^{ab},A^{abc})$. The entropy current is given by the formula
\begin{equation}
s^a = \dfrac{\partial \chi}{\partial \zeta_a } - \zeta_A N^{aA}.
\end{equation}
We compare the two states $\zeta_A$ and $\zeta_A +\delta \zeta_A$ and consider the second-order variation of the entropy current:
\begin{equation}\label{divergenceSa}
 \delta s^a := s^a[\zeta_A+\delta \zeta_A]-s^a[\zeta_A] = \dfrac{1}{2} \dfrac{\partial^3 \chi}{\partial \zeta_a \partial \zeta_A \partial \zeta_B} \delta \zeta_A \delta \zeta_B - \zeta_A \delta N^{aA}- \delta \zeta_A \delta N^{aA}.
\end{equation}
Imposing the consistency of this expression with the covariant Gibbs relation (as demanded by step i) produces the equilibrium conditions $\zeta=\alpha$, $\zeta_a =\beta K_a$ (with $\alpha,\beta=\text{const}$) and $\zeta_{ab}=0$, in agreement with \cite{GerochLindblom1990}. Combining this result with the constraint \eqref{deltaNunz}, we find that the term $- \zeta_A \delta N^{aA}$ in \eqref{divergenceSa} does not contribute to the total flux \eqref{due}, so we will use the shorthand notation
\begin{equation}\label{zNaa}
- \zeta_A \delta N^{aA} =(\text{zfc}),
\end{equation}
which stands for ``zero flux contribution''. The final step consists of using \eqref{paritacchiuhc} to write the last term in \eqref{divergenceSa} explicitly, so that
\begin{equation}
- \delta \zeta_A \delta N^{aA} = - \dfrac{\partial^3 \chi}{ \partial \zeta_a \partial \zeta_A \partial \zeta_B} \delta \zeta_A \delta \zeta_B,
\end{equation}
and we finally obtain
\begin{equation}\label{fourteen}
\delta s^a = -  \dfrac{1}{2} \dfrac{\partial^3 \chi}{\partial \zeta_a \partial \zeta_A \partial \zeta_B} \delta \zeta_A \delta \zeta_B + (\text{zfc}) = -E^a +(\text{zfc}). 
\end{equation}
The four-vector $E^a$ is the ``energy current'' introduced by \citet{GerochLindblom1990} in equation (51), but we see here that it is actually a second-order entropy current, whose flux is the difference between the entropy in equilibrium (the state defined by $\zeta_A$) and the entropy in the perturbed state (the state defined by $\zeta_A +\delta \zeta_A$):
\begin{equation}
E =-\int_\Sigma E^a \, d \Sigma_a = S_{\text{eq}} -S .
\end{equation}
Therefore, the condition of maximality of the entropy in equilibrium ($S_{\text{eq}}\geq S$) is equivalent to the positivity requirement for the ``energy functional'', $E \geq 0$, see figure \ref{fig:fig}. Note that most of the mathematical properties of the field equations (e.g. their symmetric-hyperbolicity) are irrelevant for this stability criterion, because our method is based on the constitutive relations \eqref{constutiveingeneral}. In this sense, this is a condition of \textit{thermodynamic} stability, which needs to hold independently from the dynamical equations we choose.

We have applied this same method also to the Israel-Stewart theory \cite{Israel_Stewart_1979}, obtaining an analogous result ($\delta s^a = (\text{zfc})-E^a$), where in this case the ``energy current'' coincides with the one introduced by \citet{Hishcock1983} in their stability analysis (see Supplementary Material: Part 2).
This clarifies the physical meaning of the stability conditions they obtained, showing how one can elegantly derive the energy functional $E$ from thermodynamic principles only \footnote{About their energy current, \citet{Hishcock1983} wrote: ``\textit{There was, unfortunately, no elegant derivation which one might hope to apply to other situations. Our ``derivation'' was in fact based on a series of modifications and generalizations of previously existing results for similar physical situations.}''}. Furthermore, since the theory of Eckart is a particular Israel-Stewart theory (with $\alpha_j=\beta_j=0$ \cite{Hishcock1983}) for which $E$ fails to be positive definite, we have a direct proof that the Eckart theory is unstable because the entropy is not maximised in equilibrium (in agreement with \cite{GavassinoLyapunov_2020}). An analogous argument applies to Landau-Lifshitz and, more in general, to any Fick-type diffusion law.

It is interesting to analyse an example which goes beyond the standard models of causal heat conduction and viscosity, like the case of a mixture of two chemical components (say, $p$ and $n$) which undergo a chemical reaction 
\begin{equation}\label{pinn}
p+p \ce{ <=> } p+n.
\end{equation}
This is an instructive case of study because, as we are going to see, our method treats the conditions of hydrodynamic, thermal, diffusive and chemical stability on the same footing.

If we do not model explicitly viscous effects and relative flows, as in \cite{Burrows1986,Alford2020}, the fields of the theory can be chosen to be the energy, p-particle and n-particle densities ($\rho$, $n_p$, $n_n$), plus the fluid four-velocity $u^a$, which is normalised: $u^a u_a=-1$. The constitutive relations take the perfect-fluid form \cite{noto_rel}
\begin{equation}\label{matu}
T^{ab}=(\rho+p)u^a u^b + p g^{ab}  \spc   N^a=(n_p+n_n)u^a  \spc s^a=su^a,
\end{equation}
where the equation of state depends on both particle densities,
\begin{equation}
ds = \dfrac{1}{T} d\rho - \dfrac{\mu_p}{T} dn_p - \dfrac{\mu_n}{T} dn_n \, ,
\end{equation}
and the pressure can be computed from the Euler relation
\begin{equation}
\rho + p = Ts + \mu_p n_p + \mu_n n_n.
\end{equation}
The conserved particle current $N^a$ given in \eqref{matu} is preserved by the chemical reaction \eqref{pinn}, which, on the other hand, does not conserve the currents $n_p u^a$ and $n_n u^a$ separately. Therefore, although we have two chemical species, they give rise to a single (not two) conserved charge $N$, to be held constant in the variation. 

The computation is analogous to the previous case, with the caveat that the condition $u^a u_a=-1$ needs to be respected by the variation, producing the (exact) identity $ 2 u_a \delta u^a=-\delta u^a \delta u_a$.
Taking the first-order variations and imposing the stationarity condition for $S$ produces the well-known equilibrium conditions $\mu_p/T=\text{const}$, $u^a/T = \beta K^a$ (with $\beta$ constant) and $\mu_p = \mu_n$. We focus, here, on the second-order variations, a calculation that is facilitated if one starts directly from the equilibrium state. In fact, the condition that the perturbation should preserve the energy $U$ takes the simple form $\delta T^{ab} \, u_b/T = (\text{zfc})$, which, employing the constitutive relations \eqref{matu}, can be used to prove the relation
\begin{equation}
- \dfrac{\delta(\rho u^a)}{T} = \dfrac{T^{ab} \delta u_b}{T} + \dfrac{\delta T^{ab} \delta u_b}{T} + (\text{zfc}).
\end{equation}
The second-order variation of the entropy current is
\begin{equation}
\delta s^a = \delta s \, u^a + s \, \delta u^a + \delta s \, \delta u^a,
\end{equation}
with
\begin{equation}
\delta s = \dfrac{ \delta \rho}{T} - \dfrac{\mu_p}{T} ( \delta n_p +\delta n_n)  + \dfrac{1}{2} s^{AB} \delta n_A \delta n_B, 
\end{equation}
where we have grouped the densities using the notation $n_A=(\rho,n_p,n_n)$. $s^{AB}$ are the components of the Hessian matrix of $s(n_A)$. After a bit of manipulation, combining together the above results and imposing $\delta N^a =(\text{zfc})$, we can write the second-order correction to the entropy current in terms of a quadratic ``energy current'', $\delta s^a = -E^a + (\text{zfc})$, with
\begin{equation}\label{Eachemo}
E^a = \dfrac{\delta T^a_b \delta u^b}{T} - \dfrac{u^a}{2} \dfrac{\rho + p}{T} \delta u^b \delta u_b - \dfrac{u^a}{2} s^{AB} \delta n_A \delta n_B .
\end{equation}
Following the same procedure of \citet{Hishcock1983}, one can show that imposing $E>0$ for any $\delta \varphi_i \neq 0$ is equivalent to requiring
\begin{equation}\label{ineqe}
 e := T \, \dfrac{E^a n_a}{u^b n_b}  = \dfrac{\rho+p}{2} \, \delta u^b \delta u_b- \dfrac{T}{2} \,  s^{AB} \, \delta n_A \delta n_B -\delta p \, \lambda_a \delta u^a>0,
\end{equation}
where $n^a =-n^b u_b(u^a+\lambda^a)$ is the (time-like future-directed) unit normal vector to $\Sigma$ and $\lambda^a$ is a space-like
deviation vector whose norm lies in the range $[0,1)$. The inequality \eqref{ineqe} produces a number of stability conditions of various kinds. Among them we recognise the standard conditions of hydrodynamic stability such as $\rho+p>0$ and the conditions for thermal and diffusive stability \cite{PrigoginebookModernThermodynamics2014}, like
\begin{equation}\label{duequic}
\begin{split}
& 0> s^{\rho \rho} =\dfrac{\partial^2 s}{\partial \rho^2} \bigg|_{n_p,n_n} = - \dfrac{1}{T^2 c_v} \\
& 0>  s^{pp/nn} = - \dfrac{\partial}{\partial n_{p/n}}  \bigg( \dfrac{\mu_{p/n}}{T} \bigg) \bigg|_{\rho,n_{n/p}} . \\
\end{split}
\end{equation}
We obtain also the condition of chemical stability with respect to the reaction \eqref{pinn}, namely
\begin{equation}\label{chemicstab}
\dfrac{\partial^2 s}{\partial n_p^2} \bigg|_{\rho,n_p+n_n} =
\begin{pmatrix}
 1 &  -1 \\
\end{pmatrix} 
\begin{bmatrix}
   s^{pp} & s^{pn}  \\
   s^{np} & s^{nn}  \\
\end{bmatrix} 
\begin{pmatrix}
1   \\
-1 \\
\end{pmatrix}
<0.
\end{equation}
But there are also some additional ``mixed'' conditions \footnote{To obtain \eqref{mescuzzo}, one needs to take the limit $\lambda^a \lambda_a \rightarrow 1$ and consider perturbations of the form $\delta u^a \propto \lambda^a$, $\delta n_C = \delta^A_C \delta n_A$ for a given $A$.}, such as
\begin{equation}\label{mescuzzo}
-T s^{AA} \geq \dfrac{1}{\rho + p} \, \bigg( \, \dfrac{\partial p}{\partial n_A} \bigg|_{n_B} \, \bigg)^2 \spc \forall \, A=\{ \, \rho \,, \, p \, , \,n \, \} \quad \quad (A \text{ fixed}),
\end{equation}
which cannot be derived within standard thermodynamics, nor from the perfect-fluid limit of the hydrodynamic model, but are hydro-diffusive conditions, specific of a two-component relativistic fluid. Note that, while the standard conditions of thermal and diffusive stability are necessary to guarantee that $-E^au_a \geq 0$, the ``mixed'' conditions (and all the conditions that are obtained taking $\lambda^a \lambda_a =1^-$) force $E^a$ to be time-like future-directed \cite{GavassinoCausality2021}. For this reason, \eqref{mescuzzo} is a stronger condition than \eqref{duequic}.

We also note that the existence of the reaction \eqref{pinn} leads to the condition $\mu_p=\mu_n$, but it plays no direct role in the stability criterion. This implies that, if there were no reaction, but still $\mu_p=\mu_n$ was true, we would obtain exactly the same stability conditions, but the inequality \eqref{chemicstab} would be a condition of diffusive stability. We have, thus, rediscovered the Duhem-Jougeut theorem, according to which a system that is stable to diffusion is also stable to chemical reactions \cite{PrigoginebookModernThermodynamics2014}. This is a consequence of the fact that the hydrodynamic equations (i.e. which process modifies the densities $n_p$ and $n_n$) are irrelevant, but only the constitutive relations (i.e. how the change of $n_p$ and $n_n$ affects the entropy) matter.

There is a clear similarity between \eqref{Eachemo} and the energy current of Israel-Stewart (see Supplementary Material: Part 2). Indeed, the procedure that leads to both is the same and the presence of two (or more) chemical species has essentially no practical consequence on the derivation. This implies that hypothetical extensions of Israel-Stewart to mixtures should not constitute a challenge for the computation of $E^a$. This is an important advance on conventional methods, where all the details of the hydrodynamic equations (including possible visco-chemical couplings) would need to be \textit{explicitly} accounted for.

Our method can also be applied to theories that are structurally different. If, for example, we consider a mixture of species that do not comove with each other, the structure \eqref{matu} breaks down, because a notion of fluid velocity $u^a$ does not exist out of equilibrium (there are, instead, two distinct velocities, $u_p^a$ and $u_n^a$, of respectively p-particles and n-particles). The natural formalism for describing these fluids has been formulated by Carter \cite{carter1991}. 

We have computed the ``energy current'' $E^a$ of Carter's theory in the absence of superfluidity and shear stresses \cite{Carter_starting_point}, assuming an arbitrary number of currents $n_X^a$ \footnote{The index $X$ runs over all the chemical species of the model, plus the entropy, given by $X=s$}, with conjugate momenta $\mu^X_a$, possibly in the presence of chemical reactions and relative flows. We report here only the result (for the details see Supplementary Material: Part 3),
\begin{equation}\label{TEaCarter}
TE^a = \sum_{X} \bigg[ \dfrac{u^a}{2}  \, \delta n_X^b \, \delta \mu^X_b - u^b \delta n_X^a \, \delta \mu^X_b  \bigg] \, .
\end{equation}
If all the currents comove also out of equilibrium, namely $\delta n_X^a =\delta (n_X u^a)$, \eqref{TEaCarter} reduces to \eqref{Eachemo}. However, \eqref{TEaCarter} is more general, because it is valid for completely independent variations $\delta n_X^a$ and can, therefore, be used to study the stability of a fluid against spontaneous formation of relative flows (i.e. perturbations of the form $\delta n_X^a = n_X \delta u_X^a$ with $u_a \delta u_X^a=0$ and $\delta u_X^a \neq \delta u_Y^a$).

As a last remark, we mention that there are theories in which the entropy current fails to have \textit{strictly} non-negative four-divergence, like the frame-stabilised first-order theories \cite{Kovtun2019,Shokri2020,BemficaDNDefinitivo2020}. However, this is typically the result of a first-order truncation of the entropy current. The inclusion of higher order corrections eventually restores the entropy principle \cite{Noronha2021}.

In conclusion, we have converted the hydrodynamic stability, usually regarded as a mathematical problem, into a branch of non-equilibrium thermodynamics. This fills an important gap between phenomenological hydrodynamic
modelling and statistical mechanics, providing a microscopic insight into the stability conditions of a fluid.\\

\section*{Acknowledgements}

This work was supported by the Polish National Science Centre grants SONATA BIS 2015/18/E/ST9/00577 and OPUS 2019/33/B/ST9/00942. Partial support comes from PHAROS, COST Action CA16214. The author thanks Marco Antonelli and Brynmor Haskell for reading the manuscript and providing useful comments.

\bibliography{Biblio}

\newpage

\begin{center}
\textbf{ \large Supplementary Material}
\end{center}
\riga\\

\textit{Part 1}: Using a simple thermodynamic argument, we convert the Gibbs stability criterion into the minimum grand-potential principle. This allows us to release the constraints on $\delta N$ and $\delta U$. We use this result to prove the consistency of the method with kinetic theory and statistical mechanics. \\

\textit{Part 2}: We show that the energy current of the Israel-Stewart theory, $E^a$, defined in equation (44) of \citet{Hishcock1983} is just  $- \delta s^a$, apart from a term that does not contribute to the total integral $E$. The strategy that we follow is precisely the one outlined in the main text.\\

\textit{Part 3}: Using the same technique, we compute the energy current $E^a$ of Carter's theory, with an arbitrary number of currents, in the absence of superfluidity. We show that, in the particular case of a relativistic model for heat conduction, we recover the inviscid Israel-Stewart energy current.

\newpage

\section*{Part 1: Minimum grand-potential principle}

\subsection{The method of the bath}

As explained in the main text, the Gibbs criterion demands that
\begin{equation}\label{gubbosos}
\delta S \leq 0 \spc \text{as long as} \spc \delta N = \delta U =0 \, .
\end{equation}
The presence of the constraints makes the problem of recasting $-\delta S$ into a quadratic functional harder. This is because the first-order part of $\delta S$ does not vanish for all $\delta \varphi_i$, but only for those perturbations which conserve the total energy and particle number. Luckily, there is a simple solution to this problem. 

Consider a fluid (with extensive variables $U_F,N_F,S_F$) in weak contact with an ideal heat and particle bath (with extensive variables $U_H,N_H,S_H$). The latter is defined as an effectively infinite reservoir of particles and energy with equation of state \cite{Termo,GavassinoTermometri}
\begin{equation}\label{sH}
S_H = \text{const} - \alpha \, N_H +\beta \, U_H \spc \text{with}\spc \beta \geq 0 \, ,
\end{equation}
where the parameters $\alpha$ and $\beta$ are some constants. Equation \eqref{sH} expresses the fact that the heat capacity (and, likewise, every extensive quantity) of the bath is effectively infinite. The assumption that the interaction is weak means that the extensive properties of the total system ``$\, \text{fluid} + \text{bath} \,$'' are the sum of those of the two parts:
\begin{equation}
U=U_F+U_H \spc  N=N_F+N_H \spc S=S_F+S_H \, . 
\end{equation}
The stability criterion \eqref{gubbosos} holds for the total system ``$\, \text{fluid} + \text{bath} \,$'', so that we have
\begin{equation}\label{apiub}
\delta S_F + \delta S_H \leq 0 \spc \text{as long as} \spc \delta N_F =-\delta N_H \quad \quad \delta U_F =-\delta U_H \, . 
\end{equation}
Combining \eqref{sH} with \eqref{apiub} we obtain
\begin{equation}\label{omleq}
\delta \Omega \geq 0
\end{equation}
with
\begin{equation}\label{Omega}
\Omega = U_F-\dfrac{1}{\beta} \, S_F - \dfrac{\alpha}{\beta} \, N_F \, .
\end{equation}
Equation \eqref{omleq} means that the equilibrium state of a thermodynamic system (in our case, a fluid) in contact with a heat and particle bath with (red-shifted) temperature $1/\beta$ and (red-shifted) chemical potential $\alpha/\beta$ is the state that minimizes the function $\Omega$ (with no constraint). This is nothing but the minimum grand-potential principle on curved space-time, which straightforwardly generalizes its analogue on flat space-time \cite{Callen_book,peliti_book}.

Now, let us define the functional
\begin{equation}\label{gruzzo}
E := \beta \, \delta \Omega = \beta \, \delta U_F -  \delta S_F -\alpha \, \delta N_F  \, .
\end{equation}
This functional is positive definite for all possible $\delta \varphi_i$, with or without constraints. On the other hand,
\begin{equation}\label{eight}
\begin{split}
& \text{if } \quad \delta N_F = \delta U_F =0 \spc \Rightarrow \spc E=-\delta S_F \geq 0 \\
& \text{if }\quad \delta N_F = \delta S_F =0 \spc \Rightarrow \spc E/\beta= \delta U_F \geq 0 \\
& \text{if }\quad \delta N_F  =0 \spc \spc \Rightarrow \spc E/\beta = \delta U_F - \delta S_F/\beta \geq 0 \, , \\
\end{split}
\end{equation}
which are respectively the maximum entropy principle ($S$ is maximum for fixed $N$ and $U$), the minimum energy principle ($U$ is minimum for fixed $N$ and $S$) and the minimum free energy principle ($F=U-S/\beta$ is minimum for fixed $N$). The first line of \eqref{eight} is particularly interesting for us: if we compute $E=\beta \, \delta \Omega$ in the \textit{absence} of constraints, we obtain a functional that automatically reduces to $-\delta S$ when $\delta N = \delta U=0$. However, contrarily to $-\delta S$, this functional $E$ remains positive definite also when $\delta N \, \delta U \neq 0$. Therefore, rather than computing $-\delta S$, one can directly compute $E=\beta \, \delta \Omega$ for free variations, and impose its positive definiteness. 

Indeed, the reader can verify explicitly that, in all the examples we propose (e.g. in Part 2 of this Supplementary material), the final formula for $E=-\delta S$ (written as the total flux of a current $E^a$) is exactly the same formula that one would obtain imposing $E=\beta \, \delta \Omega$ with released constraints.

 \subsection{Application 1: stability of the equilibrium in kinetic theory}

It is interesting to note that the field nature of $\varphi_i$ has never been used explicitly in the paper. This implies that the Gibbs criterion should remain valid also in the context of relativistic kinetic theory. In particular, if one replaces the fields $\varphi_i=\varphi_i(x)$ with the invariant distribution function $f=f(x,p)$, counting the number of particles in a small phase-space volume centered on $(x,p)$, all the arguments of the paper remain valid. Let us verify it explicitly. We set our units of energy in such a way that $g_s/h^3 =1$, where $g_s$ is the spin degeneracy of the gas and $h$ is Planck's constant. 

The entropy current, particle current and stress-energy tensor of an ideal quantum gas are (working in local inertial coordinates)
\begin{equation}
    \{ \, s^a \, , \, N^a \, , \, T^{ab} \, \} = \int \{ \, \mathfrak{s} \, , \, f \, , \, f p^b \, \} \, p^a \, \dfrac{ d^3 p}{p^0} \, ,  
\end{equation}
where $\mathfrak{s} = \mathfrak{s} (f)$ is a function that depends on the type of particle. Following \citet{WuAnyons1994}, a reasonably general formula for $\mathfrak{s}$ is
\begin{equation}\label{sazuz}
\mathfrak{s}= - f \ln f - (1-af)\ln(1-af) + (1+\tilde{a}f)\ln(1+\tilde{a}f) \, ,
\end{equation}
where $\tilde{a}=1-a$. Bosons have $a=0$, while Fermions have $a=1$. Intermediate cases (namely $0 < a < 1$) are anyons, which can exist in 2+1 dimensions. The Maxwell-Boltzmann case is recovered taking the limit of small $f$, with arbitrary $a$. All thermodynamic equilibrium states can be computed imposing the covariant Gibbs relation $\delta s^a = -\alpha \, \delta N^a - \beta K_b \, \delta T^{ab}$ (to first order), which implies
\begin{equation}
    \dfrac{d \mathfrak{s}}{df} = -\alpha -\beta_b p^b   \spc \beta_b = \beta K_b \, .
\end{equation}
Computing explicitly the derivative of \eqref{sazuz} we obtain the equilibrium condition
\begin{equation}
(1-af)^a (1+\tilde{a} f)^{\tilde{a}} = f e^{-\alpha - \beta_b p^b} \, ,
\end{equation}
which is the covariant generalization of the equilibrium occupation law given in \cite{WuAnyons1994}. 

We compute the functional $E$ from equation \eqref{gruzzo}. It can be written as the flux of the current $E^a=-\delta s^a -\alpha \delta N^a -\beta_b \delta T^{ab}$ which, truncated to second order, reads explicitly
\begin{equation}
    E^a = - \int \dfrac{d^2 \mathfrak{s}}{d f^2} \, \dfrac{\delta f^2}{2} \, p^a \, \dfrac{ d^3 p}{p^0}.
\end{equation}
The first-order part in $\delta f$ cancels, due to the covariant Gibbs relation. The explicit formula of the second derivative of $\mathfrak{s}$ is
\begin{equation}
 \dfrac{d^2 \mathfrak{s}}{d f^2} = - \dfrac{1}{f(1-af)(1+\tilde{a}f)} < 0 \, .
\end{equation}
The inequality follows from the fact that $0 \leq f \leq 1/a$ \cite{WuAnyons1994} and implies that $E^a$ is always time-like future-directed. Recalling that $\Sigma$ is always taken space-like, we get
\begin{equation}
E = -\int_\Sigma E^a d \Sigma_a \geq 0 \spc \forall \, \delta f \, .
\end{equation}
This proves that, in kinetic theory, all thermodynamic equilibria (both rotating and non-rotating\footnote{To have a rotating equilibrium, it is sufficient to require that $\beta_b$ is not hypersurface-orthogonal, namely $\beta_{[a} \nabla_b \beta_{c]}\neq 0$ \cite{Wald}.}) in curved space-time are maximum entropy (and minimum grand-potential) states, for all types of particles. The quantity $E$ plays the role of a bounded square-integral norm, which is always larger than 0 (whenever $\delta f \neq 0$), and can only decrease in time. In conclusion, all thermodynamic equilibria are Lyapunov-stable, as long as the H-theorem (namely $\nabla_a s^a \geq 0$ \cite{cercignani_book}) holds.

\subsection{Application 2: grand-canonical ensemble in curved space-times}

We can use equation \eqref{gruzzo} to derive the formula for the equilibrium density operator of a fluid in curved space-time from thermodynamic principles.

Let us begin by considering a well-known inequality: given two density operators $\hat{\rho}$ and $\hat{\sigma}$, it is always true that \cite{RelativeEntropy2000}
\begin{equation}\label{relativentropy}
\text{Tr}\big( \hat{\sigma} \ln \hat{\sigma} \big) \geq \text{Tr}\big(\hat{\sigma} \ln \hat{\rho} \big) \, . 
\end{equation}  
If we keep $\hat{\sigma}$ arbitrary and we choose $\hat{\rho}$ to be equal to
\begin{equation}\label{rho}
\hat{\rho} = \dfrac{e^{\alpha \hat{N}-\beta \hat{U}}}{\text{Tr} \, e^{\alpha \hat{N}-\beta \hat{U}}} \, ,
\end{equation}
where $\hat{U}$ and $\hat{N}$ are the quantum energy and particle operators, then equation \eqref{relativentropy} becomes
\begin{equation}\label{ineq}
\Omega[\hat{\sigma}] \geq \Omega[\hat{\rho}] \spc \forall \, \hat{\sigma} \, ,
\end{equation}
where 
\begin{equation}\label{ommuz}
\Omega[\hat{\sigma}] = \text{Tr}\big( \hat{\sigma} \hat{U} \big) + \dfrac{\text{Tr}(\hat{\sigma} \ln \hat{\sigma})}{\beta} -\dfrac{\alpha}{\beta} \, \text{Tr}(\hat{\sigma}\hat{N}) \, .
\end{equation}
Considering that the bridge between thermodynamics and statistical mechanics is built by making the identifications
\begin{equation}
\{ \, U \, , \, N \, , \, S \, \} = \text{Tr} \bigg( \hat{\sigma} \, \{ \, \hat{U} \, , \, \hat{N} \, , \, -\ln \hat{\sigma} \, \} \bigg) \, ,
\end{equation}
it follows that $\Omega[\hat{\sigma}]$  coincides with the functional $\Omega$ introduced in equation \eqref{Omega}. Therefore, recalling that in equation \eqref{ineq} $\hat{\sigma}$ is completely arbitrary, we can conclude, invoking the minimum grand-potential principle, that \eqref{rho} is the equilibrium density operator.

We can rewrite equation \eqref{rho} in a slightly more familiar form. Recalling that the partition function $Z$ and the inverse-temperature four-vector $\beta^b$ are given by
\begin{equation}
Z = \text{Tr} \, e^{\alpha \hat{N}-\beta \hat{U}} \spc \beta^b = \beta K^b \, ,
\end{equation} 
and considering that the operators $\hat{U}$ and $\hat{N}$ can be written as the fluxes
\begin{equation}
\{ \, \hat{U} \, , \, \hat{N} \, \} = \int_\Sigma \, \{ \, K_b \hat{T}^{ab} \, , \, -\hat{N}^a \,  \} \, d\Sigma_a \, ,
\end{equation}
equation \eqref{rho} becomes
\begin{equation}\label{gradncC}
\hat{\rho} = \dfrac{1}{Z} \exp \int_\Sigma \bigg(-\alpha \hat{N}^a - \beta_b \hat{T}^{ab}   \bigg) d\Sigma_a  \, .
\end{equation}
This is the well-known formula for the equilibrium density operator of relativistic fluids in curved space-time \cite{VANWEERT1982,BecattiniPRL2012,BecattiniQuantumCorrections2015,BecattiniBeta2016}. 

The present discussion shows that the Gibbs stability criterion, as it is formulated in the main text, is fully consistent with Zubarev's approach to relativistic statistical mechanics. This remains true also in the fully non-linear regime, considering that, for the argument above to be valid, $\hat{\sigma}-\hat{\rho}$ does not need to be small.

\newpage

\section*{Part 2: Israel-Stewart theory}
\subsection{Notation}

We recall that the signature is $(-,+,+,+)$ and $c=k_B=1$. 
We adopt exactly the same notation as \citet{Hishcock1983}, with only three differences: for us $s$ is the entropy per unit volume, $\sigma$ is the entropy per particle $(s=n\sigma)$ and the symbol $\Theta$ of \cite{Hishcock1983} is replaced by the more conventional notation $\mu/T$. This is done to guarantee coherence of notation with the main text.

\subsection{The constitutive relations of the Israel-Stewart theory}

We interpret the Israel-Stewart theory as a field theory for the tensor fields
\begin{equation}\label{IsraelFields}
(\varphi_i) = (u^a,\rho,n,\tau,q^a,\tau^{ab}),
\end{equation}
representing respectively the flow velocity, the rest-frame energy and particle densities, the bulk-viscous stress, the heat flux and the shear-viscous stress. They satisfy the algebraic constraints
\begin{equation}\label{algebrain}
u^a u_a+1 = u^a q_a = u^a \tau_{ab}=\tau_{[ab]}=\tau^a_a =0.
\end{equation}
Introducing the projector $q^{ab}=g^{ab} +u^a u^b$, the constitutive relations for the conserved fluxes are
\begin{equation}\label{ISstress}
\begin{split}
& T^{ab} = \rho u^a u^b +(p+\tau)q^{ab} + u^a q^b + u^b q^a + \tau^{ab}  \\
& N^a = nu^a \\
\end{split}
\end{equation}
and the one for the entropy current is
\begin{equation}\label{constituentropy}
s^a = su^a + \dfrac{q^a}{T} - (\beta_0 \tau^2+\beta_1 q^b q_b +\beta_2 \tau_{bc}\tau^{bc}) \dfrac{u^a}{2T} + \dfrac{\alpha_0 \tau q^a}{T} +\dfrac{\alpha_1 \tau^a_b q^b}{T},
\end{equation}
where $\alpha_j$ and $\beta_j$ are some expansion coefficients.
The quantities $s$ and $p$ (representing the equilibrium entropy density and pressure) are connected to $\rho$ and $n$ by means of the equilibrium equation of state $s=s(\rho,n)$, hence (defined the equilibrium temperature $T$ and chemical potential $\mu$) we have
\begin{equation}\label{entropon}
ds = \dfrac{1}{T} d\rho - \dfrac{\mu}{T} dn
\end{equation}
and
\begin{equation}\label{Euler}
\rho +p = Ts+\mu n.
\end{equation}

\subsection{The equilibrium states}

The fact that the equilibrium states of the Israel-Stewart theory can be computed from an entropy principle is a well-known foundational feature of the theory \cite{Israel_Stewart_1979}. Therefore, we will not perform the step (i) of the method (namely the first-order analysis) explicitly , as we already know that the equilibrium conditions that we would obtain from the requirement $\delta S=0$ (at the first order) are precisely the conditions of zero entropy production ($\nabla_a s^a = 0$) found by \citet{Hishcock1983}, namely:
\begin{equation}\label{Equil1}
\tau = q^a = \tau^{ab}=0
\end{equation}
and
\begin{equation}\label{Equil2}
\dfrac{\mu}{T} = \alpha  \spc  \dfrac{u^a}{T} = \beta K^a  \spc \alpha,\beta=\text{const}  \spc  \beta>0.
\end{equation}
We recall that the physical setting we are considering is the one outlined in our letter: stationary background spacetime, with a unique time-like future-directed symmetry generator $K^a$.

\subsection{Constraints on the second-order variations}

The whole study is based on the comparison between an equilibrium state $\varphi_i$, which obeys the conditions \eqref{Equil1}-\eqref{Equil2}, and a slightly perturbed state $\varphi_i + \delta \varphi_i$,  which models a small deviation from equilibrium. Both these states are assumed to obey the Israel-Stewart hydrodynamic equations (equations that, however, we do not need to introduce explicitly). The variations $\delta \varphi_i$ need to obey some constraints. First of all, since the algebraic constraints \eqref{algebrain} must hold for both $\varphi_i$ and $\varphi_i + \delta \varphi_i$, this produces the following exact identities:
\begin{equation}\label{identities}
u^a \delta u_a = -\dfrac{\delta u^a \delta u_a}{2}  \spc u^a \delta q_a = -\delta u^a \delta q_a  \spc u^a \delta \tau_{ab} = -\delta u^a \delta \tau_{ab}  \spc  \delta \tau_{[ab]}=\delta \tau^a_a =0 ,
\end{equation}
where we made use also of the condition \eqref{Equil1}, to be imposed on the unperturbed fields. We, furthermore, recall that the metric tensor is treated as a fixed background field, which is unaffected by the perturbation (implying that, e.g., $\delta u_a = g_{ab} \delta u^b$). The identities \eqref{identities} are very useful, because they can convert quantities which look to be of first order in the perturbation (such as $u^a \delta q_a$), into quantities that are manifestly quadratic in the variations (in our example, $-\delta u^a \delta q_a$). 

The other crucial constraints come from the requirement that $\delta N=\delta U=0$. More explicitly, we need to impose
\begin{equation}
\{ \, \delta N , \,  \delta U \, \} = \int_\Sigma \{ \, -\delta N^a, \, \delta T^{ab} K_b \, \} \, d\Sigma_a =0.
\end{equation}
Recalling \eqref{Equil2} and adopting the same notation as in the main text, we can rewrite the aforementioned constraints in the following simpler forms:
\begin{equation}\label{duesorelle}
\delta N^a =(\text{zfc})  \spc \delta T^{ab} \dfrac{u_b}{T} = (\text{zfc}).
\end{equation}
The first equation can be immediately converted into a constraint on the fields $n$ and $u^a$:
\begin{equation}\label{sorella1}
\delta n \, u^a + n \, \delta u^a + \delta n \, \delta u^a = (\text{zfc}).
\end{equation}
Furthermore, the second equation of \eqref{duesorelle} can be rewritten\footnote{Start from the general identity $\rho u^a + q^a +T^{ab} u_b=0$, valid on both the equilibrium and the perturbed state.} in the more useful form 
\begin{equation}\label{sorella2}
\dfrac{\delta \rho \, u^a}{T} + \dfrac{\rho \, \delta u^a}{T} + \dfrac{\delta \rho \, \delta u^a}{T} + \dfrac{\delta q^a}{T} +  \dfrac{T^{ab}\delta u_b}{T}  + \dfrac{\delta T^{ab} \, \delta u_b}{T} = (\text{zfc}).
\end{equation}

\subsection{Perturbation to the entropy current}

We only need to make the second-order expansion of the constitutive relation \eqref{constituentropy} in terms of $\delta \varphi_i$, where we recall that the selection of fields $\varphi_i$ to be used as free variables is made in \eqref{IsraelFields}. The calculation is straightforward:
\begin{equation}\label{dsa}
\begin{split}
\delta s^a = \, &  \bigg( \dfrac{\delta \rho}{T} -\dfrac{\mu}{T} \delta n +\dfrac{1}{2} s^{AB} \delta n_A \delta n_B \bigg)u^a + s \, \delta u^a + (\delta \rho - \mu \delta n) \dfrac{\delta u^a}{T} + \dfrac{\delta q^a}{T}       \\
& - \dfrac{\delta q^a \delta T}{T^2} - (\beta_0 \delta\tau \, \delta \tau +\beta_1 \delta q^b \delta q_b +\beta_2 \delta \tau_{bc} \delta \tau^{bc}) \dfrac{u^a}{2T} + \dfrac{\alpha_0 \delta \tau \, \delta q^a}{T} +\dfrac{\alpha_1 \delta \tau^a_b \delta  q^b}{T}, \\
\end{split}
\end{equation}
where we introduced the compact notation $n_A=(\rho,n)$ and $s^{AB}$ is the Hessian matrix of $s(n_A)$. We can use the constraints \eqref{sorella1} and \eqref{sorella2}, together with the first equilibrium condition of \eqref{Equil2} to rewrite the first line of \eqref{dsa} in a more convenient form:
\begin{equation}\label{dsa2}
\begin{split}
\delta s^a = \, &  (\text{zfc})+ \dfrac{u^a}{2} s^{AB} \delta n_A \delta n_B + (Ts +\mu n -\rho)  \dfrac{\delta u^a}{T} - \dfrac{T^{ab} \delta u_b}{T} - \dfrac{\delta T^{ab} \delta u_b}{T}   \\
& - \dfrac{\delta q^a \delta T}{T^2} - (\beta_0 \delta\tau \, \delta \tau +\beta_1 \delta q^b \delta q_b +\beta_2 \delta \tau_{bc} \delta \tau^{bc}) \dfrac{u^a}{2T} + \dfrac{\alpha_0 \delta \tau \, \delta q^a}{T} +\dfrac{\alpha_1 \delta \tau^a_b \delta  q^b}{T}. \\
\end{split}
\end{equation}
However, recalling the Euler relation \eqref{Euler}, it is easy to show that
\begin{equation}
(Ts +\mu n -\rho)  \dfrac{\delta u^a}{T} - \dfrac{T^{ab} \delta u_b}{T} = (\rho + p) \dfrac{u^a}{T}  \dfrac{\delta u^b \delta u_b}{2}, 
\end{equation}
which can be inserted into \eqref{dsa2}, giving
\begin{equation}
\delta s^a = (\text{zfc}) - E^a,
\end{equation}
with
\begin{equation}\label{energycurrent}
\begin{split}
TE^a = \, & \delta T^a_b \delta u^b - \dfrac{1}{2} (\rho +p) u^a \delta u^b \delta u_b - \dfrac{u^a}{2} T s^{AB} \delta n_A \delta n_B  \\
&   +  \dfrac{\delta q^a \delta T}{T} + (\beta_0 \delta\tau \, \delta \tau +\beta_1 \delta q^b \delta q_b +\beta_2 \delta \tau_{bc} \delta \tau^{bc}) \dfrac{u^a}{2} - \alpha_0 \delta \tau \, \delta q^a - \alpha_1 \delta \tau^a_b \delta  q^b  ,  \\
\end{split}
\end{equation}
which constitutes the quadratic ``energy current'' we were looking for.

\subsection{Comparison with the energy current of Hiscock and Lindblom}

Note that, if our task was just to compute the energy current $E^a$ of Israel-Stewart, we could just stop here. In fact, we have already obtained a formula for it: equation \eqref{energycurrent}. However, if we compare it with equation (44) of \citet{Hishcock1983}, 
\begin{equation}\label{Hiscuz}
\begin{split}
TE^a = \, & \delta T^a_b \delta u^b - \dfrac{1}{2} (\rho +p) u^a \delta u^b \delta u_b  + \frac{1}{\rho + p} \bigg( \dfrac{\partial \rho}{\partial p} \bigg|_\sigma (\delta p)^2  + \dfrac{\partial \rho}{\partial \sigma}\bigg|_{p} \dfrac{\partial p}{\partial \sigma} \bigg|_{\mu/T} (\delta \sigma)^2 \bigg) \dfrac{u^a}{2} \\
&   +  \dfrac{\delta q^a \delta T}{T} + (\beta_0 \delta\tau \, \delta \tau +\beta_1 \delta q^b \delta q_b +\beta_2 \delta \tau_{bc} \delta \tau^{bc}) \dfrac{u^a}{2} - \alpha_0 \delta \tau \, \delta q^a - \alpha_1 \delta \tau^a_b \delta  q^b  ,  \\
\end{split}
\end{equation}
we see that the two energy currents in \eqref{energycurrent} and \eqref{Hiscuz} are the same only if one manages to show that
\begin{equation}\label{maC}
 -Ts^{AB} \delta n_A \delta n_B = \frac{1}{\rho + p} \bigg( \dfrac{\partial \rho}{\partial p} \bigg|_\sigma (\delta p)^2  + \dfrac{\partial \rho}{\partial \sigma}\bigg|_{p} \dfrac{\partial p}{\partial \sigma} \bigg|_{\mu/T} (\delta \sigma)^2 \bigg)  .
\end{equation}
It turns out that this identity is, indeed, true, proving that our energy current is exactly the same as the one of \citet{Hishcock1983} and confirming the argument of \citet{GavassinoLyapunov_2020}, according to which the stability conditions of Israel-Stewart are precisely those conditions for which the entropy is maximal in equilibrium. However, proving \eqref{maC} is not so straightforward, and requires some elaborate thermodynamic manipulations, which are presented below.

First of all, we list the thermodynamic identities that are needed to prove \eqref{maC}:
\begin{equation}\label{IDDO1}
\dfrac{d T}{T} = \dfrac{d p}{\rho +p}  - \dfrac{nT}{\rho +p} \, d \bigg( \dfrac{\mu}{T} \bigg) \, , 
\end{equation}
\begin{equation}\label{IDDO15}
\dfrac{dn}{n} = \dfrac{d\rho}{\rho +p} - \dfrac{Tn \, d\sigma}{\rho +p} \, ,
\end{equation}
\begin{equation}\label{IDDO2}
\dfrac{\partial \rho}{\partial \sigma} \bigg|_p = n^2 T^2 \, \dfrac{\partial}{\partial p} \bigg( \dfrac{\mu}{T} \bigg)\bigg|_{\sigma} \, ,
\end{equation}
\begin{equation}\label{IDDO3}
  \dfrac{\partial \rho}{\partial \sigma} \bigg|_p  \dfrac{\partial p}{\partial \sigma}\bigg|_{\mu/T} = -n^2 T^2 \dfrac{\partial}{\partial \sigma}  \bigg( \dfrac{\mu}{T} \bigg)\bigg|_p\, .
\end{equation}
Equations \eqref{IDDO1} and \eqref{IDDO15} can be straightforwardly derived from the differentials $ dp = s\, dT + n \, d\mu$ and \eqref{entropon}.
Equation \eqref{IDDO2} and \eqref{IDDO3} are simply the identities (89) and (94) of \citet{Hishcock1983}.

Our proof of the identity \eqref{maC} follows four steps. First, using \eqref{entropon}, it is easy to show that (since all the terms are quadratic in the perturbation, we can use first-order identities to make changes of variables)
\begin{equation}\label{Ilprimoordine}
\mathcal{Z} := -Ts^{AB} \delta n_A \delta n_B = \delta n \, \delta \mu + \delta s \, \delta T =  T \, \delta n \, \delta \bigg( \dfrac{\mu}{T} \bigg) + \dfrac{\delta \rho \, \delta T}{T} \, .
\end{equation}
Secondly, we can use the identities \eqref{IDDO1} and \eqref{IDDO15} to justify the following equalities:
\begin{equation}
\mathcal{Z} = T \, \delta n \, \delta \bigg( \dfrac{\mu}{T} \bigg) + \dfrac{\delta \rho \, \delta p}{\rho + p} - \dfrac{nT \, \delta \rho}{\rho +p} \delta \bigg( \dfrac{\mu}{T}  \bigg) = \dfrac{\delta \rho \, \delta p}{\rho + p} - \dfrac{n^2 T^2 \delta \sigma}{\rho +p}  \delta \bigg( \dfrac{\mu}{T} \bigg) \, .
\end{equation}
The third step consists of writing $\delta \rho$ and $\delta(\mu/T)$ in terms of $\delta p$ and $\delta \sigma$,
\begin{equation}
\delta \rho = \dfrac{\partial \rho}{\partial p}\bigg|_\sigma \delta p + \dfrac{\partial \rho}{\partial \sigma}\bigg|_p \delta \sigma  \spc \delta \bigg( \dfrac{\mu}{T} \bigg) = \dfrac{\partial }{\partial p} \bigg( \dfrac{\mu}{T} \bigg)\bigg|_\sigma \delta p + \dfrac{\partial }{\partial \sigma}\bigg( \dfrac{\mu}{T} \bigg)\bigg|_p \delta \sigma \, ,
\end{equation}
so that we find
\begin{equation}
(\rho + p)\mathcal{Z} = \dfrac{\partial \rho}{\partial p}\bigg|_\sigma (\delta p)^2 + \bigg[ \dfrac{\partial \rho}{\partial \sigma}\bigg|_p - n^2 T^2 \dfrac{\partial }{\partial p} \bigg( \dfrac{\mu}{T} \bigg)\bigg|_\sigma \bigg] \delta p \, \delta \sigma - n^2 T^2 \dfrac{\partial }{\partial \sigma}\bigg( \dfrac{\mu}{T} \bigg)\bigg|_p (\delta \sigma)^2 \, .
\end{equation}
Finally, we only need to use the identities \eqref{IDDO2} and \eqref{IDDO3} to obtain
\begin{equation}
(\rho + p)\mathcal{Z} = \dfrac{\partial \rho}{\partial p}\bigg|_\sigma (\delta p)^2 +  \dfrac{\partial \rho}{\partial \sigma} \bigg|_p  \dfrac{\partial p}{\partial \sigma}\bigg|_{\mu/T} (\delta \sigma)^2,
\end{equation}
which is what we wanted to prove.

\newpage

\section*{Part 3: Carter's theory}

\subsection{Notation}

We adopt exactly the same notation as \citet{Carter_starting_point}, with the only difference that their quantities $\Psi$, $\Theta_a$ and $\Theta$ will be denoted by $p$, $T_a$ and $T$ ($p$ and $T$ reduce to the ordinary pressure and temperature in equilibrium). This is done to ensure notational conformity with Part 2. Equation \eqref{TEaCarter} of the main text is explicitly obtained in subsection \ref{quilacurr}.

\subsection{The constitutive relations of Carter's theory}

We choose the momentum-based representation, according to which the fundamental fields of the theory are the momenta
\begin{equation}
(\varphi_i) = (\mu^X_a).
\end{equation}
The theory postulates that there is a scalar field $p$ (representing the total pressure) such that the constitutive relations for the currents $n_X^a$, entropy current included (for $X=s$ we impose $n_X^a=n_s^a=s^a$), are given by the differential (at fixed metric components)
\begin{equation}\label{dpippo}
dp = -n_X^a d\mu^X_a,
\end{equation}
while the constitutive relation for the stress-energy tensor is
\begin{equation}\label{CarterStressenergy}
T\indices{^a _b} = p g\indices{^a _b} + n_X^a \mu^X_b.
\end{equation}
We are adopting Einstein's summation convention for the chemical index $X$, including $X=s$. The covector $\mu_a^s$, which is associated with the entropy current $s^a$, is denoted by $T_a$. We assume that no species is superfluid, which implies that no constraint is imposed on the covector fields $\mu_a^X$ (i.e. there is no conserved winding number \cite{Termo}).

\subsection{The equilibrium states}

Also in Carter's theory the equilibrium states can be easily computed from the maximum entropy principle. Since the calculation is straightworward, here we report only the result. Given the definition of the inverse-temperature four-vector 
\begin{equation}
\beta^a := \dfrac{-s^a}{s^b T_b},
\end{equation}
one finds that all the currents are collinear to $\beta^a$ in equilibrium (there is no superfluidity here \cite{Termo}),
\begin{equation}
\dfrac{n_X^a}{\sqrt{-n_X^b n_{Xb}}} = \dfrac{\beta^a}{\sqrt{-\beta^b \beta_b}} =:u^a  \spc \forall \, X,
\end{equation}
so that $u^a$ represents the \textit{equilibrium} collective fluid velocity of all the species. In this configuration the fluid becomes indistinguishable from a multi-constituent perfect fluid, like the $pn$-mixture presented in the main body. In equilibrium (and only in equilibrium), $T=-s^a T_a/s$ is the ordinary temperature of the mixture and we can write   
\begin{equation}\label{Equil1Carter}
\beta^a = \dfrac{u^a}{T}  \spc n_X^a = n_X u^a.
\end{equation}
Apart from the collinearity condition, which may be seen as the condition of local thermodynamic equilibrium (analogous to \eqref{Equil1} ), we have some conditions of global equilibrium (analogous to \eqref{Equil2} ):
\begin{equation}\label{Equil2Carter}
\beta^a \mu^X_a = -\alpha^X  \spc  \beta^a = \beta K^a  \spc \alpha^X ,\beta=\text{const}  \spc  \beta>0.
\end{equation}
Note that $\alpha^s=1$, identically. Finally, coherently with our remarks on the Duhem-Jougeut theorem, one can verify that the possible presence of chemical reactions does not modify any of these equilibrium conditions, but it imposes constraints on the possible values of the constants $\alpha^X$. For example, the presence of a reaction like
\begin{equation}
X +2Y \ce{ <=> } 5Z
\end{equation}
would produce the constraint
\begin{equation}\label{equilireactzio}
\alpha^X + 2 \alpha^Y = 5 \alpha^Z.
\end{equation}

\subsection{Constraints on the second-order variations}

Contrary to the case of the Israel-Stewart theory, there is no local constraint to be imposed on the variation of the fields $\mu^X_a$. All the constraints have a global character. The requirement that the variation should preserve the values of the integrals of motion produces constraints
\begin{equation}\label{CarterCoonstr}
\alpha^X \delta n_X^a = (\text{zfc})+\delta s^a  \spc \delta T\indices{^a _b} \beta^b = (\text{zfc}).
\end{equation}
While the second one is the obvious analogue of the second relation of \eqref{duesorelle}, the first one requires a bit of explanation. Let $Q_Y$ be a basis of independent conserved (i.e. unchanged by the chemical reactions) charges of the fluid,  
\begin{equation}
Q_Y = \sum_{X \neq s} q\indices{^X _Y} N_X,
\end{equation}
where $q\indices{^X _Y}$ is a matrix of constant coefficients, measuring the amount of charge $Y$ carried by an individual particle of type $X$ ($N_X$ is the total number of $X$-particles). All the equilibrium conditions of the kind   \eqref{equilireactzio} are simultaneously respected if and only if there is a set of constant coefficients $\lambda^Y$ (one for every charge $Q_Y$) such that 
\begin{equation}
\alpha^X = \sum_Y \lambda^Y q\indices{^X _Y}  \spc  \forall \, X \neq s.
\end{equation} 
Since the perturbation conserves the values of the constants of motion of the fluid, we need to impose the exact constraint $\delta Q_Y=0$, which implies
\begin{equation}
\sum_{X \neq s} \alpha^X \delta N_X = \sum_{(X\neq s),Y} \lambda^Y q\indices{^X _Y} \delta N_X = \sum_Y \lambda^Y \delta Q_Y =0,
\end{equation}  
which, written in terms of currents, becomes
\begin{equation}
\sum_{X \neq s} \alpha^X \delta n_X^a =(\text{zfc}).
\end{equation}
Adding to both sides $\delta s^a$, and recalling that $\alpha^s=1$, we finally obtain the first relation in \eqref{CarterCoonstr}.

\subsection{Perturbation to the entropy current}\label{quilacurr}

We now derive equation \eqref{TEaCarter} of the main text. Let us, first of all, consider the perturbation to the stress-energy tensor:
\begin{equation}
\delta T\indices{^a _b} = \delta p \, g\indices{^a _b} + \delta n_X^a \, \mu^X_b + n_X^a \, \delta \mu^X_b + \delta n_X^a \, \delta \mu^X_b.
\end{equation}
If we contract this variation with $\beta^b$ and impose the constraints \eqref{CarterCoonstr} we find
\begin{equation}
\delta s^a =(\text{zfc}) + \beta^a \delta p + \beta^b n_X^a \, \delta \mu^X_b + \beta^b \delta n_X^a \, \delta \mu^X_b.
\end{equation} 
The collinearity condition \eqref{Equil1Carter} implies that $\beta^b n_X^a= \beta^a n_X^b$ so we obtain
\begin{equation}
\delta s^a =(\text{zfc}) + \beta^a (\delta p +  n_X^b \, \delta \mu^X_b) + \beta^b \delta n_X^a \, \delta \mu^X_b.
\end{equation}
Finally, the second-order variation $\delta p$ can be written in the convenient form
\begin{equation}
\delta p = - n_X^b \, \delta \mu^X_b - \dfrac{1}{2} \, \delta n_X^b \, \delta \mu^X_b \, ,
\end{equation}
so that again we have $\delta s^a = (\text{zfc})-E^a$, with
\begin{equation}\label{TEaCarter2}
TE^a = \dfrac{u^a}{2}  \, \delta n_X^b \, \delta \mu^X_b - u^b \delta n_X^a \, \delta \mu^X_b \, .
\end{equation}
This Hiscock-Lindblom-type current can be used to derive all the stability conditions of a generic (non-superfluid) Carter's fluid, both in the presence and in the absence of chemical reactions. 

\subsection{A particular case: Carter's model for heat conduction}

Carter's model for heat conduction \cite{noto_rel} is built using only two covectors ($T_a$ and $\mu_a$) as fundamental fields, which are dual respectively to the entropy and the particle current ($s^a$ and $N^a$). \citet{PriouCOMPAR1991} has shown that, close to equilibrium, this model becomes very similar to an Israel-Stewart heat-conductive (but inviscid) fluid. This comparison becomes more evident if one makes the decomposition (see \citet{Lopez09} for all the details)
\begin{equation}\label{quarantenove}
N^a = n u^a  \spc s^a = su^a + \dfrac{q^a}{T}   \spc \mu_a = \mu u_a + \dfrac{\mathcal{A} \, q_a}{T} \spc T_a =Tu_a + \dfrac{\mathcal{C} \, q_a}{T} \, ,
\end{equation}
where
\begin{equation}
u^a q_a=0 \, , \spc \mathcal{C}= \beta_{IS} T^2  \spc \text{and} \spc T = \mathcal{C}s+\mathcal{A}n.
\end{equation}
Note that, within Carter's approach, the quantities $n$, $s$, $T$, $\mu$ and $u^a$ are built from the geometrical decomposition \eqref{quarantenove} directly as non-equilibrium quantities, generalizing the corresponding equilibrium fields. Contrary to the Israel-Stewart case, they are not used as identifiers of a fiducial local thermodynamic equilibrium state. In fact, $n$ and $s$ are not connected to $T$ and $\mu$ by the equilibrium equation of state. The coefficient $\beta_{IS}$ is a sort of Carter's analogue of the thermodynamic coefficient $\beta_1$ appearing in \eqref{constituentropy}.

If we insert \eqref{quarantenove} into \eqref{TEaCarter2}, truncating the result at the second order, we obtain
\begin{equation}\label{Priousz}
\begin{split}
TE^a = \, & \dfrac{u^a}{2}(\delta n \, \delta \mu + \delta s \, \delta T) + \dfrac{u^a}{2} (n\mu + sT) \, \delta u^b \delta u_b + \delta u^a (n \delta \mu +s \delta T) \\
+ \, & u^a \delta q^b \delta u_b + \dfrac{\delta q^a \delta T}{T} + \dfrac{u^a}{2} \beta_{IS} \, \delta q^b \delta q_b \, .  \\
\end{split}
\end{equation}
To facilitate the interpretation of this current, we can insert \eqref{quarantenove} into \eqref{dpippo} and \eqref{CarterStressenergy}, to obtain the exact formulas 
\begin{equation}
dp = n \, d\mu +s \, dT - \dfrac{q^a}{T} \, d \bigg( \dfrac{\mathcal{C}q_a}{T} \bigg)
\end{equation}
\begin{equation}
\rho := T^{ab} u_a u_b = n\mu+sT-p.
\end{equation}
\begin{equation}
T^{ab} = (\rho+p)u^a u^b +p g^{ab}+ u^a q^b + u^b q^a + \beta_{IS} q^a q^b.
\end{equation}
These can be easily used to show that
\begin{equation}
\delta T^a_b \delta u^b -\dfrac{u^a}{2} (\rho + p) \, \delta u^b \delta u_b = \dfrac{u^a}{2} (n\mu +sT) \, \delta u^b \delta u_b + \delta u^a (n \delta \mu +s\delta T) + u^a \delta q^b \delta u_b \, , 
\end{equation}
so that equation \eqref{Priousz} takes the more familiar form
\begin{equation}
\begin{split}
TE^a = \, & \delta T^a_b \delta u^b -\dfrac{u^a}{2} (\rho + p) \, \delta u^b \delta u_b+ \dfrac{u^a}{2}(\delta n \, \delta \mu + \delta s \, \delta T) \\
+ \, & \dfrac{\delta q^a \delta T}{T} + \dfrac{u^a}{2} \beta_{IS} \, \delta q^b \delta q_b \, .  \\
\end{split}
\end{equation} 
Recalling equation \eqref{Ilprimoordine}, we see that this formula for $E^a$ is indistinguishable from the inviscid limit ($\delta \tau=\delta \tau^{ab}=0$) of the energy current \eqref{energycurrent} of Israel-Stewart.

\label{lastpage}

\end{document}